\documentclass[twocolumn,trackchanges]{aastex63}
\hypersetup{linkcolor=red,citecolor=blue,filecolor=cyan,urlcolor=magenta}

\usepackage{enumerate}
\usepackage{amsmath}

\received{April XX, 2022}
\revised{\today}
\accepted{??? xx, 2022}
\submitjournal{ApJ}

\shorttitle{Observing SN Neutrinos. IV. {\tt SPECIAL BLEND}}
\shortauthors{Harada et al.}
\graphicspath{{./}{fig/}}

\begin{document}

\title{Observing Supernova Neutrino Light Curves with Super-Kamiokande.\\ 
IV. Development of {\tt SPECIAL BLEND}: a New Public Analysis Code for Supernova Neutrinos}

\correspondingauthor{Akira Harada}
\email{akira.harada@riken.jp}

\author[0000-0003-1409-0695]{Akira Harada}
\affiliation{Interdisciplinary Theoretical and Mathematical Sciences Program (iTHEMS), RIKEN, Wako, Saitama 351-0198, Japan}

\author[0000-0002-7443-2215]{Yudai Suwa}
\affiliation{Department of Earth Science and Astronomy, The University of Tokyo, Tokyo 153-8902, Japan}
\affiliation{Center for Gravitational Physics and Quantum Information, Yukawa Institute for Theoretical Physics, Kyoto University, Kyoto 606-8502, Japan}

\author[0000-0003-3273-946X]{Masayuki Harada}
\affiliation{Department of Physics, Okayama University, Okayama 700-8530, Japan}

\author[0000-0003-0437-8505]{Yusuke Koshio}
\affiliation{Department of Physics, Okayama University, Okayama 700-8530, Japan}
\affiliation{Kavli Institute for the Physics and Mathematics of the Universe (Kavli IPMU, WPI), Todai Institutes for Advanced Study, \\ The University of Tokyo, Kashiwa 277-8583, Japan}

\author[0000-0002-0827-9152]{Masamitsu Mori}
\affiliation{Department of Earth Science and Astronomy, The University of Tokyo, Tokyo 153-8902, Japan}
\affiliation{Division of Science, National Astronomical Observatory of Japan, 2-21-1 Osawa, Mitaka, Tokyo 181-8588, Japan}

\author[0000-0003-4408-6929]{Fumi Nakanishi}
\affiliation{Department of Physics, Okayama University, Okayama 700-8530, Japan}

\author[0000-0001-6330-1685]{Ken'ichiro Nakazato}
\affiliation{Faculty of Arts and Science, Kyushu University, Fukuoka 819-0395, Japan}

\author[0000-0002-9224-9449]{Kohsuke Sumiyoshi}
\affiliation{National Institute of Technology, Numazu College of Technology, Numazu 410-8501, Japan}

\author{Roger A. Wendell}
\affiliation{Department of Physics, Kyoto University, Kyoto 606-8502, Japan}
\affiliation{Kavli Institute for the Physics and Mathematics of the Universe (Kavli IPMU, WPI), Todai Institutes for Advanced Study, \\ The University of Tokyo, Kashiwa 277-8583, Japan}

%\collaboration{1}{(nuLC collaboration)}

\begin{abstract}

Supernova neutrinos are invaluable signals that offer information about the interior of supernovae. Because a nearby supernova can occur at any time, preparing for future supernova neutrino observation is an urgent task. For the prompt analysis of supernova neutrinos, we have developed a new analysis code, ``Supernova Parameter Estimation Code based on Insight on Analytic Late-time Burst Light curve at Earth Neutrino Detector ({\tt SPECIAL BLEND})''. This code estimates the parameters of supernova based on an analytic model of supernova neutrinos from the proto-neutron star cooling phase. For easy availability to the community, this code is public and easily runs on web environments. {\tt SPECIAL BLEND} can estimate the parameters better than the analysis pipeline we developed in the previous paper. By using {\tt SPECIAL BLEND}, we can estimate the supernova parameters within $10\%$ precision up to $\sim 20\,{\rm kpc}$ and $\sim 60\,{\rm kpc}$ (Large Magellanic Cloud contained) with Super Kamiokande and Hyper Kamiokande, respectively.

\end{abstract}

\keywords{Core-collapse supernovae -- Supernova neutrinos -- Neutrino astronomy -- Neutrino telescopes -- Neutron stars}

%%%%%%%%%%%%%%%%%%%%%%%%%%
\section{Introduction} 
\label{sec:intro}
%%%%%%%%%%%%%%%%%%%%%%%%%%
Core-collapse supernovae (CCSNe) are the explosive death of massive stars. The promising scenario for the explosion mechanism is the delayed neutrino heating mechanism. The gravitationally collapsed stellar core bounces when the central density reaches almost nuclear density, and then the bounce shock is launched with the proto-neutron star (PNS) left at the center. The bounce shock eventually stalls due to energy loss, but the energetic neutrinos emitted from the PNS energize the shock to revive, leading to the explosion. Many theoretical works have been dedicated to uncovering the mechanism \citep{2021ApJ...915...28B, 2021Natur.589...29B, 2019MNRAS.482..351V, 2020ApJ...903...82I, 2020ApJ...902..150H, 2021ApJ...906..128K, 2020MNRAS.499.4174M, 2022arXiv221112675B}. However, we have not yet reached a conclusion due to an incomplete theoretical model and a lack of observational data.

The supernova neutrino is an essential observational clue for understanding the explosion mechanism. Among the neutrinos emitted from the proto-neutron star (PNS), only about $\mathcal{O}(1\%)$ is absorbed, and the remainder escapes the star. Since neutrinos interact weakly with stellar matter, they carry information about the central region of the supernova, where they are emitted. Detecting and analyzing these neutrinos allows us to investigate the explosion mechanism. For example, the Kamiokande \citep{1987PhRvL..58.1490H} and IMB \citep{1987PhRvL..58.1494B} experiments, which are water Cerenkov neutrino detectors, detected the supernova neutrino from SN1987A in the Large Magellanic Cloud and estimated the energy budget of core-collapse supernovae that is consistent with the energy released during neutron star formation \citep{1987Sci...237.1471S}. Unfortunately, the optical observations are not as informative about the central, optically thick region, and the utility of gravitational wave signals is unproven. Supernova neutrino observation, however, has a proven track record and is a promising signal. A Galactic supernova can occur at any time, with an expected rate of once every $\sim 30$ years per galaxy, making it important to prepare for the analysis of the next supernova neutrino observation.

Several theoretical models and analysis tools for supernova neutrino have been suggested. The popular but outdated model is the Wilson model suggested by \cite{1998ApJ...496..216T}. For more modern models, many supernova simulations have been published so far \citep{2013ApJS..205....2N, 2014PhRvD..90d5032T, 2015ApJS..219...24O, 2016ApJ...821...38S, 2016NCimR..39....1M, 2018PhRvD..98l3001W, 2020PhRvD.101l3013W, 2019MNRAS.482..351V, 2020ApJ...898..139W, 2021ApJ...906..128K, 2021Natur.589...29B, 2021ApJ...911...74Z}. {\tt SNEWPY} \citep{2021JOSS....6.3772B} is a publicly available tool mediating the theoretical supernova models and analysis tools such as {\tt SNOwGLoBES} \citep{snowglobes} and {\tt sntools} \citep{2021JOSS....6.2877M}. It also incorporates the influence of vacuum/matter neutrino oscillation. {\tt SNOwGLoBES} and {\tt sntools} are the tools to calculate the detector response from the neutrino signals and to generate mock observational data, respectively. {\tt SNOwGLoBES} is utilized in many works: \cite{2021arXiv210110624S} suggested the technique to estimate the distance to CCSNe and zero-age-main-sequence mass from early neutrino signals, and \cite{2022MNRAS.512.2806N} suggested how to estimate the PNS mass from the total emitted neutrino energy.

Some works developed techniques to discriminate theoretical models from mock neutrino observation signals. With neutrino observation by the future Hyper Kamiokande, \cite{2021ApJ...916...15A} suggested a model discrimination method using the likelihood ratio; even if the distance to the supernova is unknown, theoretical models can be discriminated up to $\sim 60 {\rm kpc}$ distance within a few percent errors. \cite{2022PhRvD.105h3017O} conducted a similar work but with Super-Kamiokande-size detectors and long-term 1D theoretical models; using neutrino signals from supernovae whose distance is unknown, models can be discriminated up to $\sim 10\,{\rm kpc}$. Although the difference between the maximum distances of \cite{2021ApJ...916...15A} and \cite{2022PhRvD.105h3017O} is larger than expected from the detector mass of Hyper Kamiokande and Super Kamiokande, we should note that we can not directly compare these works because of different sets of theoretical models.

While many works focus on the early phase (up to several seconds) of the CCSN explosion, investigation of the late phase has unique advantages. In the early phase of a supernova before shock revival, complex physical processes such as convection, standing accretion shock instability, turbulence, neutrino reactions, and nuclear equations of state make it hard to predict supernova neutrino signals precisely. On the other hand, in the late phase after an explosion, or PNS cooling phase whose mass accretion almost ceases, the system becomes relatively simple. Besides, about half of the supernova neutrino detection would be from this PNS cooling phase \citep{2021PTEP.2021b3E01M}, hence neglecting this phase wastes signals. Obtaining information about the PNS mass and radius from the neutrino signals in this late phase lays a foundation to examine the more complicated early-phase signals.

We have conducted a series of investigations on the late-phase emission of supernova neutrinos. \cite{2013ApJS..205....2N} calculated the neutrino signals of several PNS cooling models up to $20\,{\rm s}$ and published Supernova Neutrino Database. \cite{2019ApJ...881..139S} extended the PNS cooling calculation to $100\,{\rm s}$ and suggested a new analysis technique, backward time analysis. The time-backward cumulative neutrino event number distribution is useful to distinguish the model. Then, \cite{2022ApJ...925...98N} conducted similar simulations using various nuclear equations of state to provide a database with a wide range of models; and they also estimated the uncertainties of the observed signals. \cite{2021PTEP.2021b3E01M} followed the progenitor core collapse, shock revival, and PNS cooling seamlessly under spherical symmetry using a modified version of {\tt GR1D} code \cite{2015ApJS..219...24O} and a light progenitor with a steep density structure. They also suggested an integrated framework to apply such simulations for supernova neutrino analysis.

Relying upon only the numerical simulations is inefficient when analyzing observational data. All of the models referred to so far are based on numerical simulations and cover a possible ``parameter space'' (e.g., PNS mass, radius, and nuclear equation of state) only discretely. When we use these models for the analysis, we need to find the model that produces the most similar neutrino signal to the observation, and then we would conduct simulations with slightly different parameters to find the best-fit model. However, numerical simulations require high computational costs and long times.

To speed up the analysis, we suggested an analysis pipeline based on an analytic model of a supernova neutrino light curve to make a rough estimate of the parameters before the detailed numerical simulation search. The PNS cooling phase is further divided into three sub-phases: the mantle contraction phase, shallow decay phase, and volume cooling phase \citep[see, e.g.,][]{2021PhRvD.103b3016L, 2020ApJ...891..156N}. After the shallow decay phase, the PNS mass and radius are almost identical to those of the final neutron star, and the neutrino signals are easy to detect thanks to the long-lasting emission. We constructed the analytic model of neutrino emission from this shallow decay phase \citep{2021PTEP.2021a3E01S}. In this model, we calculated the neutrino emission from the PNS surface evolving quasi-stationary based on the diffusion approximation. The parameters of this analytic model are the PNS mass, radius, and the total emitted neutrino energy during the shallow decay phase. Furthermore, we developed the analysis pipeline to determine the parameters of this analytic model from the supernova neutrino observation data at a water Cerenkov detector like Super-Kamiokande \citep[ hereafter paper III]{2022ApJ...934...15S}. After getting a rough estimate of the parameters with this pipeline, we could efficiently search numerical models around them in detail efficiently.

However, we have left some points to improve in paper III. First, the analysis method is based on so-called chi-square fitting. Exactly speaking, the chi-square fitting is justified when the data obeys the Gauss distribution, while the count data, like neutrino detection, obeys the Poisson distribution. This is not a problem when analyzing a Galactic supernova because the Poisson distribution can be approximated by the Gauss distribution when the expectation value is large enough, thanks to the central limit theorem. However, it may be a problem when analyzing a distant supernova with a small number of neutrino detection. Second, we only use the mean energy of the positron produced by the inverse beta decay in the detector, but exploiting the whole energy distribution of the date reduces the loss of information.

Therefore, in this paper, we have developed a new public analysis pipeline named ``Supernova Parameter Estimation Code based on Insight on Analytic Late-time Burst Light curve at Earth Neutrino Detector ({\tt SPECIAL BLEND})'' by improving the pipeline in paper III. This pipeline estimates the parameters of the analytic model, assuming that the observed data obeys the Poisson distribution whose expectation values are calculated from the analytic model as a function of time and positron energy. With this treatment, the pipeline is improved from that in paper III. Besides, this pipeline serves as a pre-analysis of detailed analysis on numerical simulations. Thus, we developed the pipeline as a public code seen in {\tt github}\footnote{\url{https://github.com/akira-harada/SPECIAL_BLEND}} so that everyone in the community can use it before the analysis using their own models. This pipeline works not only on personal computers but also on web environments like {\tt Google Colaboratory}. The user-friendly interface is one of the advantages of {\tt SPECIAL BLEND}.

This paper is structured as follows. First, we explain the design of {\tt SPECIAL BLEND} in section \ref{sec:method}. The parameter estimation method is presented in section \ref{sec:likelihood}, and the usage of {\tt SPECIAL BLEND} is outlined in section \ref{sec:specialblend}. Then, section \ref{sec:perform} demonstrates the performance of {\tt SPECIAL BLEND}. Section \ref{sec:perform_galactic} highlights the improvement from paper III, and how the precision of the parameter estimation depends on the distance to the supernova follows in section \ref{sec:perform_distance}. Finally, we give a summary and conclusion of this paper in section \ref{sec:summary}.

%%%%%%%%%%%%%%%%%%%%%%%%%%
\section{Analysis Method and Code Description}
\label{sec:method}
%%%%%%%%%%%%%%%%%%%%%%%%%%
{\tt SPECIAL BLEND} is based on the Bayesian parameter estimation with the analytic model of the neutrino light curve and is publicly available with an easy-to-use interface. Section \ref{sec:likelihood} reviews the Bayesian analysis, derives the likelihood functions, and describes the analytic model. Afterward, the structure and usage of {\tt SPECIAL BLEND} are explained in section \ref{sec:specialblend}.

\subsection{Likelihood Analysis}
\label{sec:likelihood}
For the parameter estimation, {\tt SPECIAL BLEND} utilizes the likelihood function. Given that experimental data $X$ obeys a probability distribution function (PDF) $P(X|\theta)$ where $\theta$ is a model parameter set, the likelihood function $L$ is defined as $L(\theta|\{X_i\}) = \prod_i P(X_i|\theta)$, where $\{X_i\}$ is the set of the experimental data, and the index $i$ runs over the data set. Here, we assume that each data in the data set is statistically independent. In other words, the likelihood function is the PDF in which the roles of constant parameters and variables are exchanged. When one just wants to estimate the best-fit parameter, a usual way is to take the maximum likelihood approach, where the argument that maximizes $L$ gives the best fit.

In order to estimate the uncertainty of the parameters, {\tt SPECIAL BLEND} follows the Bayesian approach. In the Bayesian approach, the parameter is considered to follow the posterior PDF, $P(\theta|\{X_i\}) = L(\theta|\{X_i\})w(\theta)/\int d\theta L(\theta|\{X_i\})w(\theta)$, where $w(\theta)$ is the prior PDF of the parameter $\theta$. By assuming the uniform prior PDF, the posterior PDF of the parameter is the normalized likelihood function $\mathcal{L}(\theta|\{X_i\})$\footnote{The symbol $\mathcal{L}$ is always normalized, while $L$ is not necessarily normalized in this paper.}. Thus, we loosely call the parameter PDF ``normalized likelihood (function)'' instead of ``posterior PDF'' in the following of this paper, even though the prior PDF is not always uniform. With this parameter PDF, the $100(1-\alpha)\%$-credible interval (CI) is defined as the parameter region $V$ such that $1-\alpha = \int_V \mathcal{L}(\theta|\{X_i\})d\theta$ and $\mathcal{L}(\theta|\{X_i\})$ is constant on the boundary of $V$. Usually, $68\%$- and $95\%$-CIs are considered as the uncertainty of the parameter. In the following, we try to estimate the PNS mass $M$, PNS radius $R$, and total emitted neutrino energy $E$, and they are collectively denoted as $\theta=(M,R,E)$.

The least $\chi^2$ approach employed in paper III is the Bayesian approach with the Gaussian PDF. The event rate and average energy analysis essentially uses the Gaussian PDF of
\begin{eqnarray}
&&P(\{N_i, \langle \epsilon \rangle_i\}|\theta) = \nonumber \\ && \prod_i  \frac{1}{\sqrt{2\pi\mathcal{R}_i(\theta)\Delta t_i}} \exp\left(-\frac{(N_i -\mathcal{R}_i(\theta)\Delta t_i)^2}{2\mathcal{R}_i(\theta)\Delta t_i}\right)\times \nonumber \\&& \frac{1}{\sqrt{2\pi(0.05E_{{\rm e^+}i}(\theta))^2}}\exp\left(-\frac{(\langle \epsilon \rangle_i - E_{{\rm e^+}i}(\theta))^2}{2(0.05 E_{{\rm e^+}i}(\theta))^2}\right),
\end{eqnarray}
where $i$ is the time index, $\Delta t_i$ is the width of the $i$-th time bin, $N_i$, $\langle \epsilon \rangle_i$, $\mathcal{R}(\theta)_i$, and $E_{\rm e^+}(\theta)$ are the detection event number, the average detection energy, the theoretical event rate, and the theoretical average energy, respectively,  at $i$-th time bin. The theoretical quantities $\mathcal{R}$ and $E_{\rm e^+}$ are defined in equations (55) and (56) of \cite{2021PTEP.2021a3E01S}, respectively. The prior is $w(\theta) \propto \prod_i (0.05E_{{\rm e^+}i}(\theta)) \sqrt{\mathcal{R}_i(\theta)}$ whose support is the range of the parameter search. Therefore, the parameter PDF becomes
\begin{eqnarray}
&&\mathcal{L}_{\rm GA}(\theta|\{N_i, \langle \epsilon \rangle_i\}) \propto \nonumber \\ && \prod_i  \exp\left(-\frac{(r_i -\mathcal{R}_i(\theta))^2\Delta t_i}{2\mathcal{R}_i(\theta)}-\frac{(\langle \epsilon \rangle_i - E_{{\rm e^+}i}(\theta))^2}{2(0.05 E_{{\rm e^+}i}(\theta))^2}\right), \label{eq:GALH}
\end{eqnarray}
where $r_i = N_i/\Delta t_i$ is the observed count rate at $i$-th time bin. Finally, this equation with $\Delta t_i$ replaced with $N_i/\mathcal{R}_i$ is the parameter PDF evaluated in Paper III. We checked that this replacement hardly affects the parameter estimation.

Though the Gaussian PDF provides reasonable parameter estimation, the Poisson distribution is more appropriate. This is because the counting data obeys the Poisson distribution. The count rate part of the above PDF is actually the Gaussian approximation of the Poisson distribution in the sense that the mean and variance are equal. Therefore, we call the parameter PDF of equation (\ref{eq:GALH}) ``Gaussian-approximation likelihood (GALH)'' for simplicity, though the prior is not uniform in this case.

Our new analysis scheme is completely based on the Poisson distribution. In order to take energy spectral information into account, we consider two-dimensional (2D) bins of time and energy, and the index $i$ runs over all of the 2D bins. The likelihood function with the Poisson distribution is given by \citep{1989NYASA.571..601L, 1989A&A...224...49J,1996PhRvD..54.1194J}
\begin{eqnarray}
    &&\mathcal{L}_{\rm B}(\theta|\{n_i\}_{\text{2D bin}})\propto\nonumber \\
    &&\prod_{i\in \text{2D bin}} \exp\left(- \mathcal{R}_{\rm spe}(t_i,\epsilon_i,\theta) \Delta\epsilon_i \Delta t_i  \right)\times \nonumber \\ &&\frac{(\mathcal{R}_{\rm spe}(t_i,\epsilon_i,\theta)\Delta \epsilon_i \Delta t_i)^{n_i}}{n_i!},
\end{eqnarray}
where $n_i$ and $\mathcal{R}_{\rm spe}(t_i,\epsilon_i,\theta)\Delta \epsilon_i \Delta t_i$ are the observed and expected event numbers in the $i$-th 2D bin, respectively, and $\Delta t_i$ and $\Delta \epsilon_i$ are the widths of the time and energy bin, respectively. The prior is chosen to be uniform on the support of the parameter search range. The discussion about the choice of the prior is presented in Appendix \ref{sec:prior}.

The spectral count rate $\mathcal{R}_{\rm spe}$ is defined as follows:
\begin{eqnarray}
    &&\mathcal{R}_{\rm spe}(t,\epsilon,\theta) = \mathcal{R}(t,\theta)\int p(t,\epsilon_{\rm e^+},\theta)\eta(\epsilon_{\rm e^+})u(\epsilon,\epsilon_{\rm e^+})d \epsilon_{\rm e^+}, \nonumber \\ && \\
    &&\mathcal{R}(t,\theta) = 720\,{\rm s^{-1}}\left(\frac{M_{\rm det}}{32.5\,{\rm kton}}\right)\left(\frac{D}{10\,{\rm kpc}}\right)^{-2}\times\nonumber \\&&\left(\frac{M}{1.4\,M_\odot}\right)^{15/2}\left(\frac{R}{10\,{\rm km}}\right)^{-8}\left(\frac{g\beta}{3}\right)\left(\frac{t+t_0}{100\,{\rm s}}\right)^{-15/2}, \label{eq:anarate} \\
    && p(t,\epsilon,\theta) = \frac{1}{F_4 T(t,\theta)^5}\frac{\epsilon^4}{1+\exp(\epsilon/ T(t,\theta))}, \label{eq:anaspec}\\
    && T(t, \theta) = \frac{F_4}{F_5} E_{\rm e^+} \nonumber\\
    && = 25.3\,{\rm MeV} \frac{F_4}{F_5} \left(\frac{M}{1.4\,M_\odot}\right)^{3/2}\left(\frac{R}{10\,{\rm km}}\right)^{-2}\times \nonumber \\&&\left(\frac{g\beta}{3}\right)\left(\frac{t+t_0}{100\,{\rm s}}\right)^{-3/2}, \label{eq:anatemp}\\
    && t_0 = 210\,{\rm s}\left(\frac{M}{1.4\,M_\odot}\right)^{6/5}\left(\frac{R}{10\,{\rm km}}\right)^{-6/5}\times\nonumber \\&&\left(\frac{g\beta}{3}\right)^{4/5}\left(\frac{E}{10^{52}\,{\rm erg}}\right)^{-1/5},
\end{eqnarray}
where $p$ is the positron spectrum, $\eta$ is the detector efficiency, $u(\epsilon,\epsilon_{\rm e^+})$ is the probability density that the positron of energy $\epsilon_{\rm e^+}$ is detected with the energy of $\epsilon$,  $M_{\rm det}$ is the detector mass, $D$ is the distance to the supernova, $g$ is the surface structure correction factor, $\beta$ is the opacity boost factor, $t_0$ is the characteristic time scale of the neutrino emission, $T$ is the surface temperature of PNS, and $F_4$ and $F_5$ are the fourth and fifth Fermi--Dirac integral with zero chemical potential, respectively. Note that equations (\ref{eq:anarate}) and (\ref{eq:anatemp}) are the same as equations (55) and (56) of \cite{2021PTEP.2021a3E01S} and are displayed here for completeness. Because we are focusing on the late phase of supernova signals when the luminosity and spectrum of all six species of neutrinos are almost the same, we need not consider the effects of neutrino oscillation. The positron energy spectrum $p$ is derived as follows: the coming neutrinos obey the Fermi--Dirac distribution with vanishing chemical potential,
\begin{equation}
    f(t,\epsilon, \theta) = \frac{1}{1+\exp(\epsilon/T(t,\theta))},
\end{equation}
and by considering a factor of $\epsilon^2$ from the energy-space volume element $\epsilon^2 d\epsilon$ and another factor of $\epsilon^2$ from the inverse beta decay cross section $\propto \epsilon^2$, in total a factor of $\epsilon^4$, the positron energy spectrum becomes
\begin{equation}
    p(t,\epsilon,\theta) \propto \frac{\epsilon^4}{1+\exp(\epsilon/T(t,\theta))}.
\end{equation}
The denominator $F_4 T^5$ is the normalization factor. We employ several simplifications throughout this paper: we neglect the difference between the positron and neutrino energies by the reaction $Q$-value; we assume that $u(\epsilon,\epsilon_{\rm e^+}) = \delta(\epsilon-\epsilon_{\rm e^+})$ (negligible detection error) and $\eta = 1$ ($100\%$ efficiency); we neglect the background. Therefore, the likelihood function becomes
\begin{eqnarray}
    &&\mathcal{L}_{\rm B}(\theta|\{n_i\}_{\text{2D bin}})\propto\nonumber \\
    &&\prod_{i\in \text{2D bin}} \exp\left(- \mathcal{R}(t_i,\theta)p(t_i,\epsilon_i,\theta) \Delta\epsilon_i \Delta t_i  \right)\times \nonumber \\ &&\frac{(\mathcal{R}(t_i,\theta)p(t_i,\epsilon_i,\theta)\Delta \epsilon_i \Delta t_i)^{n_i}}{n_i!}. \label{eq:BLH}
\end{eqnarray}
We call this likelihood ``binned likelihood (BLH)''.

When the event number is small, BLH is inefficient. In that case, the representative time and energy of the bin $(t_i,\epsilon_i)$ differs from the event time and energy, and this error significantly degrades the data quality. If many 2D bins do not have any events, the product operation $\prod_{i\in\text{2D bin}}$ results in the multiplication of many unity factors and waste computational time. Hence, we also employ the unbinned likelihood (UBLH) analysis. The likelihood function is derived from equation (\ref{eq:BLH}) by considering the case that $\Delta t_i$ and $\Delta \epsilon_i$ are sufficiently small so that each bin never has two or more events. Then, only bins that contain an event have non-unity contributions to the product $\prod_{i\in\text{2D bin}}$. The exponential factor results in the exponential of the integral. In summary, the UBLH function is
\begin{eqnarray}
    &&\mathcal{L}_{\rm UB}(\theta|\{t_i,\epsilon_i\}_{\rm events})\nonumber \\
    &&\propto\exp\left(-\int \mathcal{R}(t,\theta)p(t,\epsilon,\theta)dt d\epsilon \right)\prod_{i\in {\rm events}}\mathcal{R}(t_i,\theta)p(t_i,\epsilon_i,\theta) \nonumber\\
    &&=\exp\left(-\int \mathcal{R}(t,\theta)dt  \right)\prod_{i\in {\rm events}}\mathcal{R}(t_i,\theta)p(t_i,\epsilon_i,\theta). \label{eq:UBLH}
\end{eqnarray}
Note that the positron spectrum is normalized to unity, and we neglect the integration element $dt_id\epsilon_i$ because it is independent of $\theta$.

\subsection{{\tt SPECIAL BLEND}}
\label{sec:specialblend}
{\tt SPECIAL BLEND} takes the observational data $\{t_i,\epsilon_i\}$ and parameters $M_{\rm det}$, $D$, $g\beta$ as input. In general, observational data of supernova neutrinos consists of neutrino emissions from several phases. Because we focus on the shallow decay phase, we must ignore the neutrinos emitted in the other phases. For this purpose, {\tt SPECIAL BLEND} cuts away the event data from the early phase (the explosion and mantle contraction phases) and the final phase (the volume cooling phase). However, the effectiveness of this function will be discussed in the forthcoming paper since this paper checks the performance of {\tt SPECIAL BLEND} by utilizing the mock observational data based on the analytic model. As for the parameters, we assume the distance to the supernova $D$ and the phenomenological correction factor $g\beta$ are determined by the optical observations and detailed theoretical modeling of the PNS, respectively, for simplicity.

The output of {\tt SPECIAL BLEND} is the normalized likelihood. The estimated parameter dimension is three; hence we span a 3D grid in parameter space of $(M,R,E)$ and directly calculate the value of the likelihood function on each grid point instead of taking the Markov-chain Monte-Carlo approach. The output is the likelihood marginalized to 2D (on $M$--$R$, $R$--$E$, and $E$--$M$) and 1D (on $M$, $R$, and $E$), where marginalization is to integrate out variables out of the focus. For 2D marginalized likelihood, the code also indicates the best-fit parameter where the 2D likelihood is maximum and the level of the edge of the $68/95\%$-CIs. For 1D marginalized likelihood, the code shows the best fit (maximum of the 1D likelihood) and $68/95\%$-CIs directly.

{\tt SPECIAL BLEND} comprises the main Fortran file and interface files. The main file contains subroutines binning data, calculating the likelihood, and marginalizing. For the interface to run {\tt SPECIAL BLEND}, we prepare the python and Fortran versions. The python interface is an {\tt Jupyter} notebook file intended to run with the software {\tt f2py}. The user can easily run the python interface in a web environment such as {\tt Google Colaboratory}\footnote{https://colab.research.google.com/}. Thanks to the notebook format, the python interface shows the color maps of 2D marginalized likelihoods and their best fit and CIs by contours. It also outputs the graph of the 1D marginalized likelihood. The Fortran interface requires another plotting software because it outputs only numerical files, but computational time is usually shorter than the python interface on {\tt Google Colaboratory}.

{\tt SPECIAL BLEND} has three modes corresponding to the three likelihoods, equations (\ref{eq:UBLH}, UBLH), (\ref{eq:BLH}, BLH), and (\ref{eq:GALH}, GALH). All these modes are available both in the Fortran and python interfaces. Furthermore, we attached the python code to generate mock data based on the analytic solution. In this code, the total event number is drawn from the Poisson distribution whose expectation value is $\int_0^\infty dt \mathcal{R}(t,\theta)$, and then the event time and energy are assigned to each event according to equations (\ref{eq:anarate}) and (\ref{eq:anaspec}). This code is independent of {\tt SPECIAL BLEND}, but the python interface can call this. For the Fortran interface, the user must independently run the generator code and {\tt SPECIAL BLEND}.

In order to use {\tt SPECIAL BLEND}, access the {\tt github} (\url{https://github.com/akira-harada/SPECIAL_BLEND}) website. By following the instruction presented in the manual in the repository, you will estimate the supernova parameters with mock observational data or your own observational data in the future.

\section{Performance}
\label{sec:perform}
In order to evaluate the performance of {\tt SPECIAL BLEND}, we conducted several test analyses based on the analytic model. The procedure is as follows: giving a certain parameter set of the analytic model, generating mock observational data, analyzing the data by {\tt SPECIAL BLEND}, and checking that the resultant parameter estimation is consistent with the given parameter set. In section \ref{sec:perform_galactic}, we show the test result for a supernova at Galactic center to grasp the basic feature of {\tt SPECIAL BLEND}. Then, in section \ref{sec:perform_distance}, the distance (or the event number) dependence of the performance of {\tt SPECIAL BLEND} is investigated.

\subsection{Parameter estimation for a supernova at Galactic center}
\label{sec:perform_galactic}
\begin{figure}[tbp]
\centering
\includegraphics[width=0.45\textwidth]{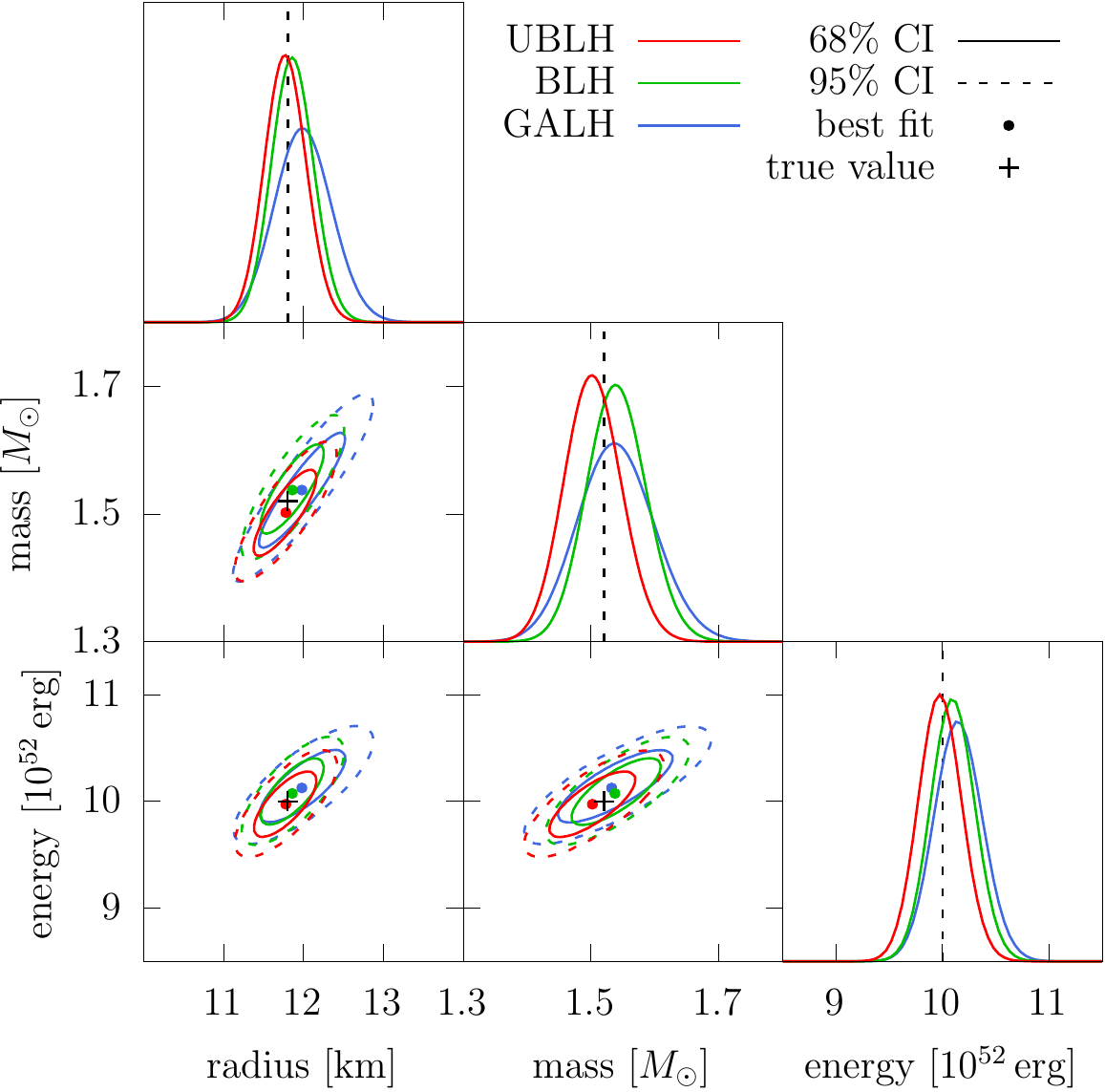}
\caption{Triangle plot for the mass, radius, and energy estimation. For the 2D likelihood, the best-fit values and $68/95\%$ CIs are shown. The different colors indicate different likelihoods: red--UBLH, green--BLH, and blue--GALH. Black crosses and vertical dashed lines indicate the true values.}
\label{fig:1real}
\end{figure}

Figure \ref{fig:1real} shows an example of the parameter estimation results using UBLH, BLH, and GALH. The ``true'' parameters to generate the mock observation data is $M=1.52\,M_\odot$, $R=11.8\,{\rm km}$, $E=1.00 \times 10^{53}\,{\rm erg}$, $g\beta = 1.6$, $M_{\rm det}=32.5\,{\rm kton}$, and $D = 8\,{\rm kpc}$. When running {\tt SPECIAL BLEND}, we fixed $M_{\rm det}$, $g\beta$, $D$, and the time origin of the data as the true values. For BLH and GALH, we set the number of the time bin $N_{\rm time}=20$. The number of energy bin $N_{\rm energy}$ for BLH is chosen to $30$. The figure shows the 1D marginalized likelihoods at diagonal positions and the best fit and $68/95\%$ CIs of 2D marginalized likelihoods at off-diagonal positions. The true values are also indicated. Apparently, all likelihoods estimate the parameters fairly well; UBLH and BLH have higher peaks than GALH, indicating smaller uncertainties. In this sense, UBLH and BLH, based on the Poisson distribution, are better than GALH.

\begin{figure}[tbp]
\centering
\includegraphics[width=0.45\textwidth]{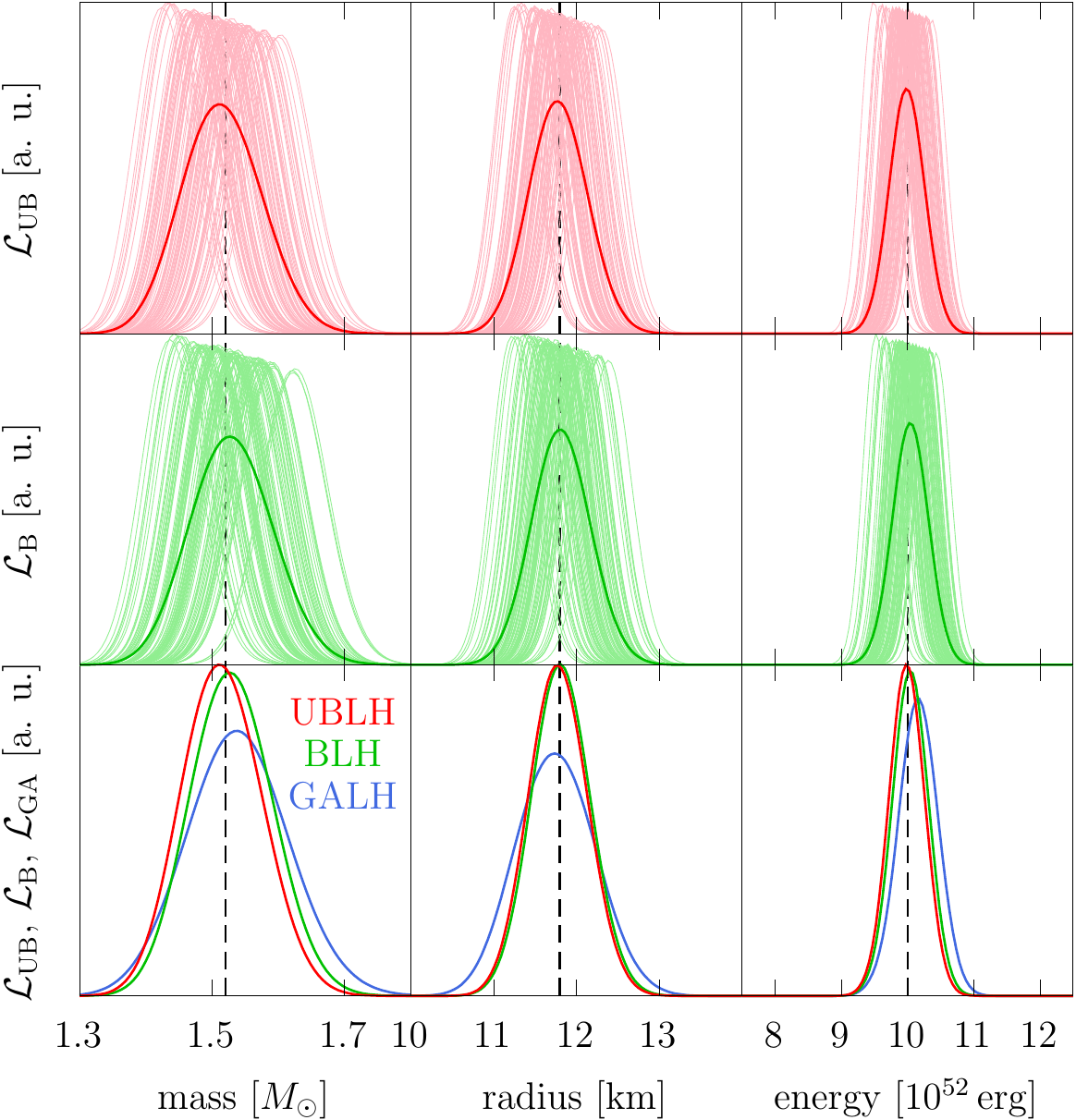}
\caption{1D marginalized likelihoods for different realizations of mock observational data. The top (UBLH) and middle (BLH) rows show the parameter estimation results of all realizations (thin lines with lighter colors) and the averages of them (thick lines with darker colors). The bottom row compares the averages of the UBLH, BLH, and GALH. For each panel, true values are indicated by the vertical dashed lines.}
\label{fig:ensamble}
\end{figure}

In order to check how well {\tt SPECIAL BLEND} works on average, we generated $100$ realizations of mock observational data with the same parameters as in figure \ref{fig:1real}, $M=1.52\,M_\odot$, $R=11.8\,{\rm km}$, $E=1.00 \times 10^{53}\,{\rm erg}$, $g\beta = 1.6$, $M_{\rm det}=32.5\,{\rm kton}$, and $D = 8\,{\rm kpc}$. Figure \ref{fig:ensamble} shows the results of the parameter estimation using all realizations and their averages. Note that we only show UBLH and BLH because a similar result for GALH is indicated in Figure 4 of \cite{2021PTEP.2021a3E01S}. As seen in the figure, the parameter estimation results have some random scatter, and hence the best fit and CIs do not always capture the true values. However, their averages do capture the true values. The average PDFs of UBLH and BLH are similar, and GALH has a broader distribution than the other two; hence, again, UBLH and BLH are better than GALH. Table \ref{tab:CI} indicates the number of realizations whose $68/95\%$ CIs by BLH contain true values out of $100$ realizations. This experiment verifies $68/95\%$ CIs.

\begin{table}
\centering
\caption{The number of realizations whose BLH CI contains the true value\label{tab:CI}}
\begin{tabular}{ccc}
\tableline
& $68\%$ CI & $95\%$ CI \\
\tableline
\tableline
mass & 64 & 97 \\
radius & 67 & 97 \\
energy & 74 & 98 \\
\tableline
\end{tabular}
\end{table}

\subsection{Distance dependence of the parameter estimation precision}
\label{sec:perform_distance}
As we have confirmed the performance of {\tt SPECIAL BLEND} for supernovae at Galactic center, we explore the distance dependence of the parameter estimation precision. For this purpose, we generated many realizations of mock observational data for supernovae at several distances. Here, the parameters are the same as in the previous section except for the distance: $M=1.52\,M_\odot$, $R=11.8\,{\rm km}$, $E=1.00 \times 10^{53}\,{\rm erg}$, $g\beta = 1.6$, $M_{\rm det}=32.5\,{\rm kton}$, and $D = 8$, $20$, $40$, $60$, $80$, $100$, $150$, $200$, $250\,{\rm kpc}$. For each $D$, we again generated $100$ realizations. Because the number of parameters to estimate is three, we discard the supernova signals with a total neutrino event count of less than three; we do not re-run the event generator and hence the number of realizations analysed in the following is less than $100$ for large $D$ cases. Note that with these parameters, the total expected count of neutrino events is $3.05$ at $D=250\,{\rm kpc}$, and hence we do not consider further distances.

Because the BLH and UBLH result in a similar performance, which to use depends on the computational time. The computational times roughly scale with the number of neutrino events $N_{\rm events}$ for UBLH and $N_{\rm time}\times N_{\rm energy}$ for BLH, respectively; hence if $N_{\rm events}$ is large, using BLH saves the computational time. For example, if we consider $N_{\rm time} = 20$ and $N_{\rm energy} = 30$, which are the default values of {\tt SPECIAL BLEND}, it is recommended to use UBLH for $N_{\rm events} \la 300$ while BLH for the other case. Indeed, when using {\tt SPECIAL BLEND} in {\tt Google Colaboratory}, the computational time to calculate UBLH $T_{\rm comp, UBLH}$ is $T_{\rm comp, UBLH} = N_{\rm events} \times 0.253\,{\rm s} + 2.12\,{\rm s}$, while the BLH counterpart $T_{\rm comp,BLH}$ is $T_{\rm comp,BLH} = 86\,{\rm s}$ for $N_{\rm time} = 20$ and $N_{\rm energy} = 30$, leading to $N_{\rm events} \sim 330$ for $T_{\rm comp,UBLH} = T_{\rm comp,BLH}$.

\begin{figure*}[tbp]
\centering
\includegraphics[width=\textwidth]{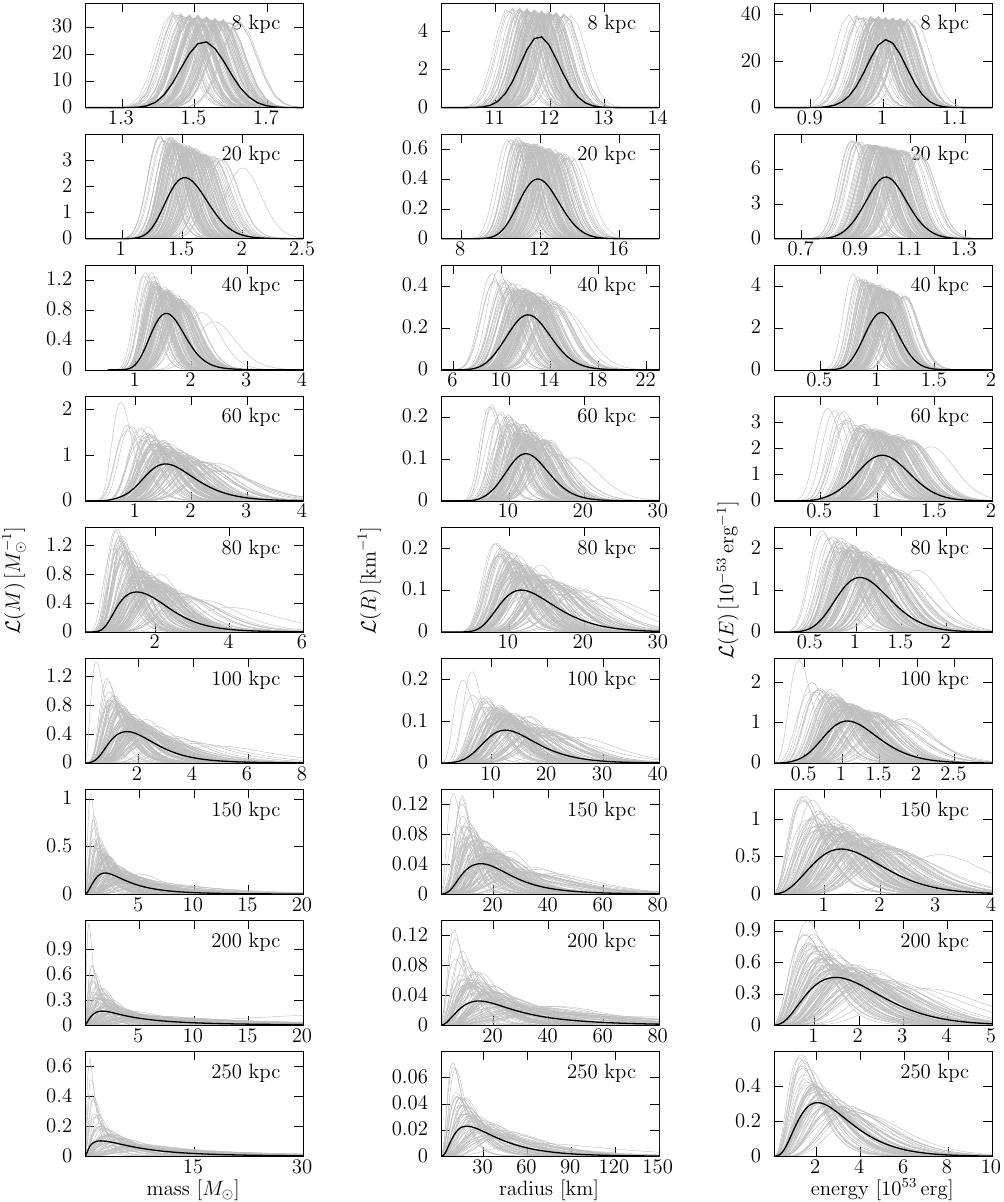}
\caption{The parameter PDFs at various distances. PDFs for each realization are indicated by thin gray lines, while the thick black line shows the average. The left, middle, and right columns correspond to the mass, radius, and energy, respectively. The distances are indicated at the upper-right corner of each panel; they are in ascending order from top to bottom. The analysis mode, BLH or UBLH, is selected according to the neutrino event number; see text for detail.}
\label{fig:ensembledist}
\end{figure*}

Figure \ref{fig:ensembledist} shows the parameter estimation results for all the realizations and their averages at different distances. We switch UBLH and BLH according to $N_{\rm events}$. The smaller distances are, the more minor the errors are. The likelihoods have tails at larger parameter values for distant supernovae with few neutrino events, while they are more symmetric for near supernovae. This behavior originates from the functional form of the likelihood. The detailed discussion is presented in appendix \ref{sec:tail}.

\begin{figure}[htb]
\centering
\includegraphics[width=0.4\textwidth]{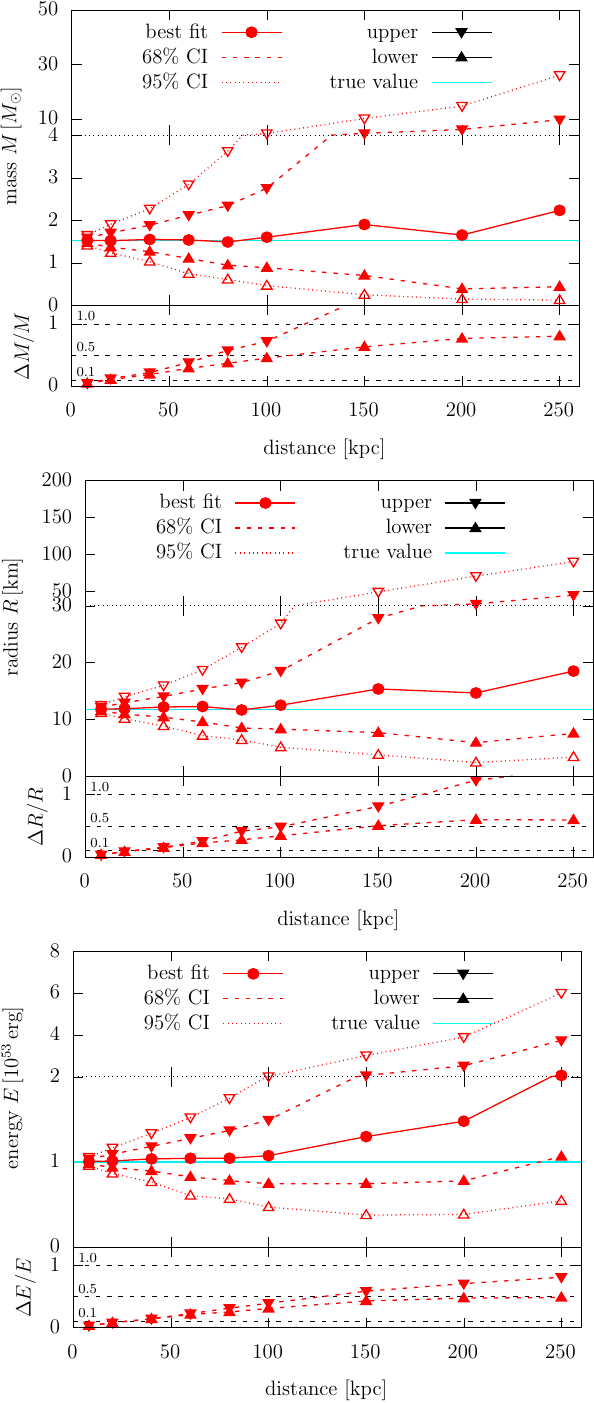}
\caption{Distance dependence of parameter best fits and $68/95\%$ CIs. The average PDFs of $100$ realizations are used to determine best fits and CIs. The top, middle, and bottom panels correspond to the mass, radius, and energy, respectively. The upper and lower portions for each panel indicate the parameter ranges and $68\%$ relative uncertainties, respectively. The true values are indicated by cyan lines.}
\label{fig:CIdist}
\end{figure}

By collecting the results of Figure \ref{fig:ensembledist}, we show the peaks and $68/95\%$ CIs of the average PDFs as functions of distance in Figure \ref{fig:CIdist}. We also indicate the $68\%$ uncertainties divided by the best fits (relative uncertainties) for each parameter. According to the figure, the best-fit values trace the true values reasonably well up to $100\,{\rm kpc}$. For further distances, the best-fit values tend to be larger, as expected from the asymmetric distribution seen in Figure \ref{fig:ensembledist}.

With the relative uncertainties shown in Figure \ref{fig:CIdist}, we see the maximum distance up to which one can estimate the parameters within given uncertainty tolerances. The parameters of supernovae within $\sim 20\,{\rm kpc}$, $\sim 70\,{\rm kpc}$, and $\sim 120\,{\rm kpc}$ can be determined within $10\%$, $50\%$, and $100\%$ uncertainties, respectively. These limits are provided from the mass estimation, which is the most stringent. The maximum distances for all parameters are indicated in table \ref{tab:maxdistanceSK}. It is noteworthy that the energy can always be determined within $100\%$ uncertainty. Although these maximum distances depend on the true parameters, we suffice with a rough estimate based on a single parameter set.

\begin{table}[htb]
\centering
\caption{The Super-Kamiokande maximum distances up to which the parameters can be determined within given uncertainty tolerances. \label{tab:maxdistanceSK}}
\begin{tabular}{cccc}
\tableline
tolerance & mass & radius & energy\\
\tableline
\tableline
$10\%$ & $\sim 20\,{\rm kpc}$ & $\sim 25\,{\rm kpc}$ & $\sim 25\,{\rm kpc}$ \\
$50\%$ & $\sim 70\,{\rm kpc}$ & $\sim 100\,{\rm kpc}$ & $\sim 130\,{\rm kpc}$ \\
$100\%$ & $\sim 120\,{\rm kpc}$ & $\sim 170\,{\rm kpc}$ & - \\
\tableline
\end{tabular}
\end{table}

\begin{table}[htb]
\centering
\caption{The same as table \ref{tab:maxdistanceSK} but for Hyper-Kamiokande. \label{tab:maxdistanceHK}}
\begin{tabular}{cccc}
\tableline
tolerance & mass & radius & energy \\
\tableline
\tableline
$10\%$ & $\sim 60\,{\rm kpc}$ & $\sim 80\,{\rm kpc}$ & $\sim 80\,{\rm kpc}$ \\
$50\%$ & $\sim 210\,{\rm kpc}$ & $\sim 300\,{\rm kpc}$ & $\sim 400\,{\rm kpc}$ \\
$100\%$ & $\sim 360\,{\rm kpc}$ & $\sim 510\,{\rm kpc}$ & - \\
\tableline
\end{tabular}
\end{table}

{\tt SPECIAL BLEND} can determine the parameters of a Galactic supernova within $10\%$ uncertainty. Thus, this public code will be useful for analyzing the future detection of supernova neutrinos. Besides, Hyper-Kamiokande is scheduled to start in 2027. Its detector volume is $260\,{\rm kton}$, $8$ times larger than that of Super-Kamiokande \citep{2018arXiv180504163H}. With equation (\ref{eq:anarate}), the detector volume increase by a factor of $8$ is equivalent to the maximum distance increase by a factor of $\sqrt{8}=2.8$. Therefore, using Hyper-Kamiokande, a supernova at $\la 60\,{\rm kpc}$ can be used for the parameter estimation within $10\%$ uncertainty. The maximum distances for Hyper-Kamiokande are shown in table \ref{tab:maxdistanceHK}. Because the distance to the LMC is $\sim 50\,{\rm kpc}$, we could estimate the parameters of SN1987A within $10\%$ uncertainty if Hyper-Kamiokande were built before 1987.

\section{Summary and Conclusion}
\label{sec:summary}
We have developed the public code for supernova neutrino analysis, {\tt SPECIAL BLEND}. This code estimates the mass, radius, and total energy of emitted neutrinos from the PNS cooling phase. This is based on the Bayesian analysis to estimate the supernova parameters. We offer three options of {\tt SPECIAL BLEND} with the different probability distributions of neutrino event rate and energy: GALH that employs the approximation of the Gaussian distribution, BLH that uses Poisson distribution and time- and energy-binned data, and UBLH that uses Poisson distribution again but time and energy of each event. We offer the Fortran and python interfaces that work on personal computers and web environments such as {\tt Google Colaboratory}.

We then examined the performance of {\tt SPECIAL BLEND}. We generated an ensemble of mock data based on the analytic model and analyzed them to check if we could reproduce the true parameters. UBLH and BLH showed almost equally better performance than GALH. If the number of neutrino events is large (small), BLH (UBLH) is better in computational time. Next, we checked the dependence of the distances to the supernova. The parameter PDFs are symmetric for nearby supernovae, while they have long tails to large parameter values for distant supernovae. When supernovae occur at $\sim 20\,{\rm kpc}$, $\sim 70\,{\rm kpc}$, and $\sim 120\,{\rm kpc}$ and Super Kamiokande detects their neutrinos, we can estimate the parameters with $10\%$, $50\%$, and $100\%$ precision, respectively. If Hyper Kamiokande will detect them, the distance becomes $2.8$ times larger with the same parameter estimation precision.

By analyzing supernova neutrinos with {\tt SPECIAL BLEND}, we would obtain a rough estimate of the parameters, and they can be utilized for further detailed analysis. However, there are several caveats. {\tt SPECIAL BLEND} is based on the analytic model, which ignores the spectral pinching of neutrinos \citep{2003ApJ...590..971K}. In order to incorporate the pinching parameter, we need a more sophisticated analytic model. Besides, the analytic model focuses on the shallow decay phase, while actual data would contain signals from the other phases. For this purpose, {\tt SPECIAL BLEND} is designed to use the events in a given time interval. It is expected that the signals from the shallow decay phase are extracted by choosing the interval appropriately. Though determining the time origin is always a problem of the supernova neutrino analysis, cutting away the events at the early phase reduces the significance of the time origin. Therefore, it is reasonable to determine the time origin as the time of the first event. However, the effectiveness of this ansatz will be presented in the forthcoming paper because it is beyond the scope of the analytic model. We are improving the model and usage of {\tt SPECIAL BLEND} and will report them elsewhere in the future.

\acknowledgments
The authors thank Catherine Beauchemin for the fruitful discussion about the Bayesian analysis. This work is supported by Grant-in-Aid for Scientific Research (20H00174, 20H01904, 20H01905, 20K03973), for Early-Career Scientists (21K13913) from Japan Society for the Promotion of Science (JSPS) and Grant-in-Aid for Scientific Research on Innovative Areas (17H06357, 17H06365, 18H05437, 19H05811, 20H04747, 22H04571) from the Ministry of Education, Culture, Sports, Science and Technology (MEXT), Japan.
For providing high performance computing resources, Computing Research Center in KEK is acknowledged.
This work was partly supported by MEXT as ``Program for Promoting Researches on the Supercomputer Fugaku'' (Toward a unified view of the universe: from large scale structures to planets). This work was partially carried out by the joint research program of the Institute for Cosmic Ray Research (ICRR), The University of Tokyo.

\appendix
\section{Presence of the tail in likelihood for distant supernovae}
\label{sec:tail}
In this appendix, we discuss the functional form of the likelihoods. Figure \ref{fig:ensembledist} shows the distance dependence of the form of the likelihood. The likelihoods with distant supernovae have the tails to larger parameter values. In other words, the likelihood has a tail if the event number is small, while it is symmetric if the event number is large. In the following, we explain the reason for the tailed form.

First, let us write down the parameter dependence of the likelihood in a simple situation. The UBLH function is defined in equation (\ref{eq:UBLH}). By directly writing down the parameter $(M,R,E)$ dependence,
\begin{eqnarray}
    \mathcal{L}_{\rm UB}(\theta|\{t_i,\epsilon_i\}_{\rm events}) &=& \exp(-N_0 m^{-3/10}r^{-1/5}e^{13/10})\prod_i \frac{\mathcal{R}_0\epsilon_i^4}{F_4T_0^5}r^2\frac{1}{1+\exp(\epsilon_i T_0^{-1}m^{-3/2}r^2(\tilde{t_i}+\tilde{t_0})^{3/2})} \nonumber \\
    &=& \exp(-N_0 m^{-3/10}r^{-1/5}e^{13/10})\prod_i \frac{\mathcal{R}_0\epsilon_i^4}{F_4T_0^5}r^2\frac{1}{1+\exp(\epsilon_i T_0^{-1}m^{-3/2}r^2(\tilde{t_i}+\tau m^{6/5}r^{-6/5}e^{-1/5})^{3/2})},\nonumber \\&&
\end{eqnarray}
where the parameters are normalized as $m=M/1.4\,M_\odot$, $r=R/10\,{\rm km}$, $e=E/10^{52}\,{\rm erg}$, $\tilde{t}_i = t_i/100\,{\rm s}$, and $\tilde{t_0}=t_0/100\,{\rm s}$. We also choose $M_{\rm det}=32.5\,{\rm kton}$, $D=10\,{\rm kpc}$, and $g\beta = 3$. The expected total event number when $m=r=e=1$ is $N_0=89$. The other constants are defined through
\begin{eqnarray}
    \mathcal{R}&=&\mathcal{R}_0 m^{15/2}r^{-8}(\tilde{t}+\tilde{t_0})^{-15/2}, \\
    T &=& T_0m^{3/2}r^{-2}(\tilde{t}+\tilde{t_0})^{-3/2},\\
    \tilde{t_0} &=& \tau m^{6/5}r^{-6/5}e^{-1/5}.
\end{eqnarray}
Here, we try to get a rough picture of the functional form of the UBLH. To this end, we consider the case that the number of events is $N_0$, and all events have the same event time and positron energy $(t,\epsilon)$. The resultant likelihood is
\begin{eqnarray}
    \mathcal{L}_{\rm UB}(\theta|\{t,\epsilon\}_{\rm events}) &=& \left(\frac{\mathcal{R}_0\epsilon^4}{F_4 T_0^5}\right)^{N_0} r^{2N_0} \exp\left( - N_0 m^{-3/10} r^{-1/5} e^{13/10}\right) \times \nonumber \\&& \left\{\frac{1}{1 + \exp\left(\frac{\epsilon}{T_0}m^{-3/2}r^2(\tilde{t} + \tau m^{6/5}r^{-6/5}e^{-1/5})^{3/2} \right)}\right\}^{N_0}. \label{eq:multipleLH}
\end{eqnarray}
For later convenience, we also define a ``single'' likelihood as
\begin{eqnarray}
    \mathcal{L}_{\rm UB,single}(\theta|\{t,\epsilon\}_{\rm events}) &=& \{\mathcal{L}_{\rm UB}(\{t,\epsilon\}_{\rm events}|\theta)\}^{1/N_0} \nonumber \\
    &=& \left(\frac{\mathcal{R}_0\epsilon^4}{F_4 T_0^5}\right) r^{2} \exp\left( - m^{-3/10} r^{-1/5} e^{13/10}\right) \times \nonumber \\&& \frac{1}{1 + \exp\left(\frac{\epsilon}{T_0}m^{-3/2}r^2(\tilde{t} + \tau m^{6/5}r^{-6/5}e^{-1/5})^{3/2} \right)}. \label{eq:singleLH}
\end{eqnarray}
This single likelihood mimics the contribution from a single event to the likelihood. Besides, we assume $\exp(\epsilon/T) \gg 1$ because the relation between the mean positron energy and the temperature is $\langle \epsilon \rangle = E_{\rm e^+} = T F_5/F_4 \simeq 5.06 T$. With these assumptions,
\begin{equation}
    \mathcal{L}_{\rm UB}(\theta|\{t,\epsilon\}_{\rm events}) = \left(\frac{\mathcal{R}_0\epsilon^4}{F_4 T_0^5}\right)^{N_0} \exp\left( - N_0 m^{-3/10} r^{-1/5} e^{13/10} - N_0\frac{\epsilon}{T_0}m^{-3/2}r^2(\tilde{t} + \tau m^{6/5}r^{-6/5}e^{-1/5})^{3/2} + 2N_0 \ln r \right). \label{eq:simpleLH}
\end{equation}

\begin{figure}[htb]
\centering
\includegraphics[width=\textwidth]{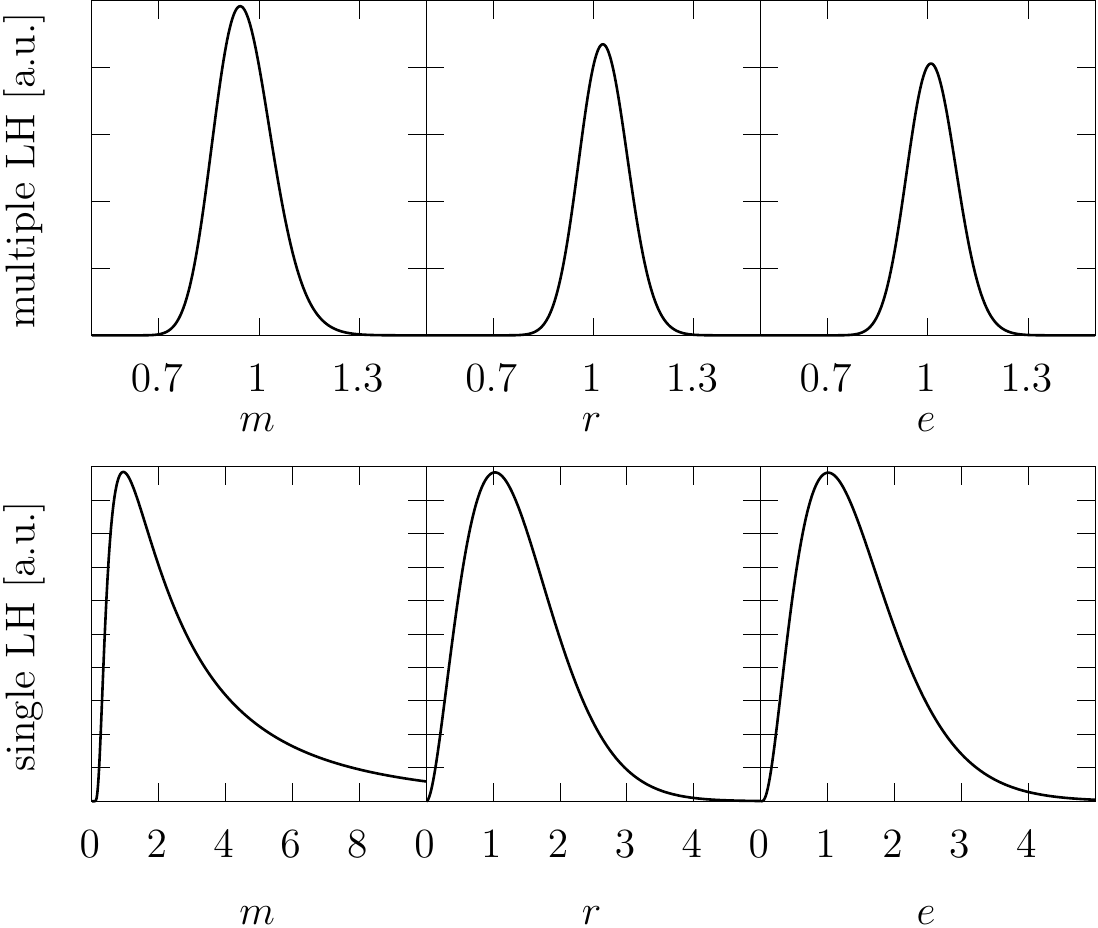}
\caption{The simplified likelihood to see the symmetrization. The lower panels show the single LH defined in equation (\ref{eq:singleLH}), while the upper panels display the likelihood in equation (\ref{eq:multipleLH}). The left, middle, and right panels are the likelihoods as functions of the normalized mass $m$, radius $r$, and energy $e$, respectively. To draw this graph, the event time and energy are fixed as $(t,\epsilon) = (30\,{\rm s}, 6.8\,{\rm MeV})$. The event energy is the mean positron energy at the event time.}
\label{fig:funcform}
\end{figure}

This simplified likelihood tells us that the likelihood is naturally asymmetric, but the product over many events symmetrizes the likelihood. To see this, we discuss mass dependence by setting $r=e=1$. For $m\ll 1$, the exponent of equation (\ref{eq:simpleLH}) becomes
\begin{equation}
    -N_0 m^{-3/10} - N_0\frac{\epsilon}{T_0}m^{-3/2} \tilde{t}^{3/2} \sim - N_0\frac{\epsilon}{T_0}\tilde{t}^{3/2}m^{-3/2}.
\end{equation}
On the other hand, when $m \gg 1$, the exponent is
\begin{equation}
    -N_0 m^{-3/10} - N_0\frac{\epsilon}{T_0}\tau^{3/2} m^{3/10} \sim - N_0\frac{\epsilon}{T_0}\tau^{3/2} m^{3/10}.
\end{equation}
Therefore, we get
\begin{equation}
    \mathcal{L}_{\rm UB}(m|\{t,\epsilon\}_{\rm events}) = \left(\frac{\mathcal{R}_0\epsilon^4}{F_4 T_0^5}\right)^{N_0}\times \left\{\begin{array}{cc} \exp\left( - N_0\frac{\epsilon}{T_0}\tilde{t}^{3/2}m^{-3/2}\right) & (m\ll1) \\  \exp\left(- N_0\frac{\epsilon}{T_0}\tau^{3/2} m^{3/10}\right) & (m\gg1)\end{array} \right..
\end{equation}
We visualize the single likelihood (\ref{eq:singleLH}) as a function of $m$ in the lower left panel of figure \ref{fig:funcform}. If the number of events is small, this likelihood shows the tail owing to the different mass dependence for the two limiting cases. However, if the number of events is large, the product of the exponential factor over whole events suppresses the tail compared to the peak (the upper left panel of figure \ref{fig:funcform} shows the likelihood (\ref{eq:multipleLH}) for $N_0$ events), resulting in a more like symmetric likelihood. Similarly, we consider the radius and energy dependencies with $m=e=1$ and $m=r=1$, respectively. The limiting values of the exponent are
\begin{equation}
    \text{the exponent}=\left\{\begin{array}{cc}
    -N_0 r^{-1/5} - N_0 \frac{\epsilon}{T_0}\tau^{3/2} r^{1/5} + 2N_0 \ln r \sim -N_0 r^{-1/5} & (r\ll1)\\
    -N_0 r^{-1/5} - N_0 \frac{\epsilon}{T_0}\tilde{t}^{3/2} r^2 + 2N_0 \ln r \sim - N_0 \frac{\epsilon}{T_0}\tilde{t}^{3/2} r^2& (r\gg1)
    \end{array}
    \right.
\end{equation}
for radius dependence and
\begin{equation}
    \text{the exponent}=\left\{\begin{array}{cc}
    -N_0 e^{13/10} - N_0 \frac{\epsilon}{T_0}\tau^{3/2} e^{-3/10} \sim - N_0 \frac{\epsilon}{T_0}\tau^{3/2} e^{-3/10} & (e\ll1)\\
    -N_0 e^{13/10} - N_0 \frac{\epsilon}{T_0}\tilde{t}^{3/2} \sim -N_0 e^{13/10}& (e\gg1)
    \end{array}
    \right.
\end{equation}
for energy dependence.
These exponents result in
\begin{equation}
    \mathcal{L}_{\rm UB}(r|\{t,\epsilon\}_{\rm events}) = \left(\frac{\mathcal{R}_0\epsilon^4}{F_4 T_0^5}\right)^{N_0} r^{2N_0}\times \left\{\begin{array}{cc} \exp\left( -N_0 r^{-1/5}\right) & (r\ll1) \\  \exp\left(- N_0 \frac{\epsilon}{T_0}\tilde{t}^{3/2} r^2\right) & (r\gg1)\end{array} \right.,
\end{equation}
and
\begin{equation}
    \mathcal{L}_{\rm UB}(e|\{t,\epsilon\}_{\rm events}) = \left(\frac{\mathcal{R}_0\epsilon^4}{F_4 T_0^5}\right)^{N_0} \times \left\{\begin{array}{cc} \exp\left( - N_0 \frac{\epsilon}{T_0}\tau^{3/2} e^{-3/10}\right) & (e\ll1) \\  \exp\left(-N_0 e^{13/10}\right) & (e\gg1)\end{array} \right..
\end{equation}
Again, these likelihoods have tails for a small number of events (the lower middle and right panels of figure \ref{fig:funcform} show the single likelihood (\ref{eq:singleLH})), and the products over a large number of events result in the suppression of the tail and symmetric likelihood (the upper middle and right panels of figure \ref{fig:funcform} show the likelihood (\ref{eq:multipleLH})). Although the discussion resorts to the simplified UBLH, we expect the likelihoods investigated in section \ref{sec:perform_distance} to have the same property.

\section{The dependence on the prior probability distribution}
\label{sec:prior}
\begin{figure}[htb]
\centering
\includegraphics[width=0.4\textwidth]{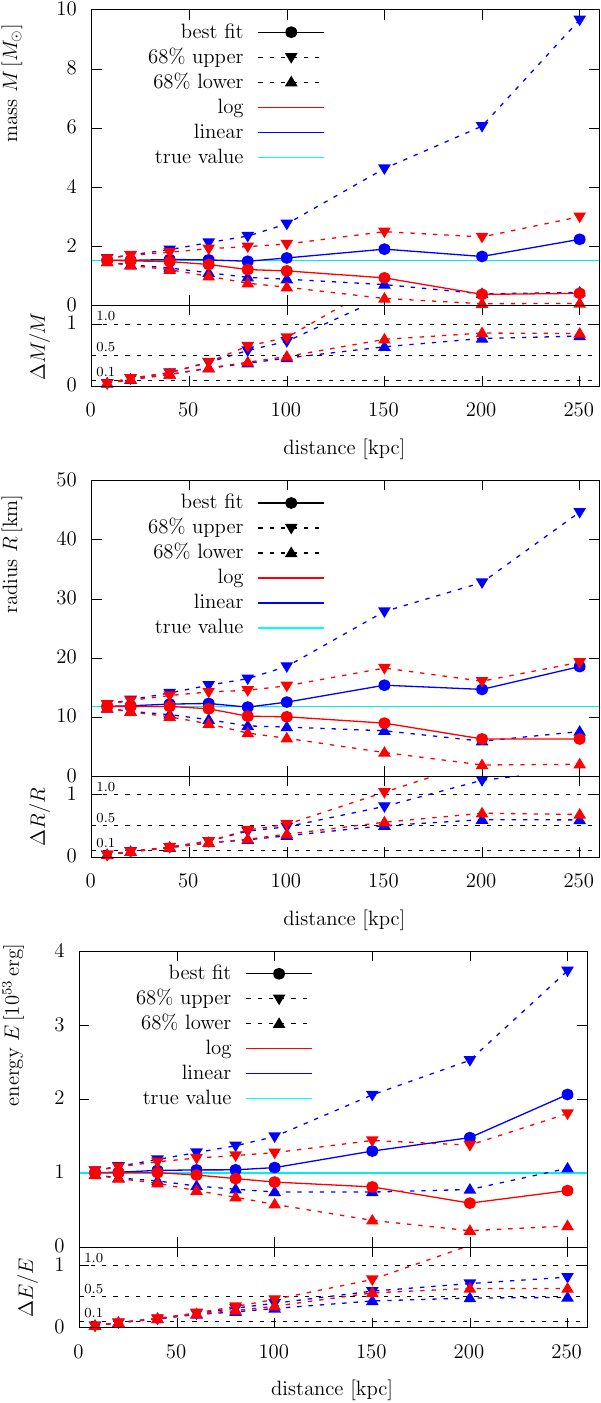}
\caption{Distance dependence of the parameter estimation similar to figure \ref{fig:CIdist}. The differences are that the red and blue lines correspond to the results using $w_{\rm log}(\theta)$ and $w_{\rm lin}(\theta)$, respectively, and the $95\%$ CIs are omitted.}
\label{fig:loglin}
\end{figure}

The choice of the prior is one of the essential factors of the Bayesian approach. We employed the linear-uniform prior,
\begin{equation}
    w_{\rm lin}(\theta) = \frac{dM}{M_{\rm max}-M_{\rm min}}\frac{dR}{R_{\rm max}-R_{\rm min}}\frac{dE}{E_{\rm max}-E_{\rm min}},
\end{equation}
in this paper.
On the other hand, another popular choice of the prior is the log-uniform prior,
\begin{eqnarray}
    w_{\rm log}(\theta) &=& \frac{d\ln M}{\ln M_{\rm max}-\ln M_{\rm min}}\frac{d\ln R}{\ln R_{\rm max}-\ln R_{\rm min}}\frac{d\ln E}{\ln E_{\rm max}-\ln E_{\rm min}} \nonumber \\ &=&\frac{dM}{M \ln(M_{\rm max}/M_{\rm min})}\frac{dR}{R \ln(R_{\rm max}/R_{\rm min})}\frac{dE}{E\ln(E_{\rm max}/E_{\rm min})}.
\end{eqnarray}
Figure \ref{fig:loglin} indicates the influence of the prior choice. This is similar to figure \ref{fig:CIdist} in the main text, but the best-fit values and $68\%$ CIs using $w_{\rm lin}(\theta)$ and $w_{\rm log}(\theta)$ are shown. The effect of the prior is negligible for the nearby supernovae that are in the main scope of {\tt SPECIAL BLEND}. For the distant supernovae, the results using the log-uniform prior deviate from the true values earlier than the linear-uniform prior. In this sense, the linear-uniform prior gives better estimations. That is the reason why we choose the linear-uniform prior.

\bibliography{ref.bbl}{}
\bibliographystyle{aasjournal}

%\listofchanges

\end{document}